\documentclass[twocolumn]{aa}
\usepackage{graphicx}
\usepackage{txfonts}
\begin{document}
%

\newcommand{\Y}{\vec{Y}}
\newcommand{\X}{\vec{X}}
\newcommand{\Off}{\vec{OFF}}
\newcommand{\A}{\vec{A}}
\newcommand{\B}{\vec{B}}
\newcommand{\Nn}{\vec{N}}
\newcommand{\Cc}{\vec{C}}
\newcommand{\Nos}{\vec{N}_{\rm obs}}
\newcommand{\Wos}{\vec{W}_{\rm obs}}
\newcommand{\Npix}{\vec{N}_{\rm pix}}
\newcommand{\Wpix}{\vec{W}_{\rm pix}}
\newcommand{\Q}{\mbox{$\mathrm{Q}$}}
\newcommand{\R}{\mbox{$\mathrm{R}$}}
\newcommand{\U}{\mbox{$\mathrm{U}$}}

  \def\etal{{et al.\/}}
  \def\PRD#1 {{PRD}}

   \title{An iterative destriping technique for diffuse background polarization data}

   \author{C. Sbarra\inst{1}, E. Carretti\inst{1}, S. Cortiglioni\inst{1}, 
        M. Zannoni\inst{2}, R. Fabbri\inst{3}, 
        C. Macculi\inst{1} \and M. Tucci\inst{4}}

   \offprints{C. Sbarra, \email{sbarra@bo.iasf.cnr.it}}

   \institute{IASF-CNR Bologna, via P. Gobetti 101, I-40129 Bologna, Italy\\
              \email{sbarra@bo.iasf.cnr.it, carretti@bo.iasf.cnr.it, 
             cortiglioni@bo.iasf.cnr.it, \\ macculi@bo.iasf.cnr.it}
	    \and
              IASF-CNR Milano, Via Bassini 15, I-20133 Milano, Italy\\
              \email{zannoni@mi.iasf.cnr.it}
	    \and
              Dipartimento di Fisica, Universit\`a di Firenze, via Sansone 1, 
              I-50019 Sesto Fiorentino, Italy\\
              \email{fabbrir@unifi.it}
	    \and
              Instituto de Fisica de Cantabria, Avda. Los Castros s/n, 
              39005 Santander, Spain\\
              \email{tucci@ifca.unican.es}
            }

   \date{Received 18 December 2002 / Accepted 13 February 2003}

   \abstract{We describe a simple but effective iterative procedure
     specifically designed to destripe $Q$ and $U$
     Stokes parameter data as those collected
     by the SPOrt experiment onboard the International Space Station (ISS).
     The method is general enough to be useful for other experiments, both
     in polarization and total intensity.  
     The only requirement for the algorithm to work properly is that the 
     receiver knee frequency must be lower than the 
     signal modulation frequency, corresponding in our case 
     to the ISS orbit  period. 
     Detailed performances of the technique are presented in the
     context of the SPOrt experiment, both in terms of added rms noise 
     and residual correlated noise. 

   }

\authorrunning{C. Sbarra et al.}
   \maketitle
%

\section{introduction}
Low frequency noise is known to affect all radiometers and to 
induce correlations among successive samples of the measured signal,
leading  to typical striping effects when producing sky maps.
It is characterised by a power-law
spectrum  $S(f)\propto(1/f)^\beta$, with $\beta$ roughly in 
the range $1 \div 2.5$ 
depending on the noise
source. The total instrumental 
noise power spectrum is usually specified in terms of 
the knee frequency, $f_{\rm k}$, at which the white and 
low frequency components of 
the noise are equal:
\begin{equation}
S(f)=\sigma_0^2\left[1+\left(\frac{f_{\rm k}}{f}\right)^{\beta}\right]
\label{eq:noiseps}
\end{equation}
where $\sigma_0^2$ represents the white noise variance
per second. As shown 
by Janssen et al. (1996), 
when data are taken from  spinning spacecrafts, where the spin acts 
as an ideal switch or chopper, most of the low frequency noise can 
be removed by software, provided $f_{\rm k}$ is lower than   the 
satellite spin frequency, $f_{\rm s}$.\\
For present and future space experiments primarily dedicated to 
Cosmic Microwave Background (CMB) anysotropy 
measurements, like MAP\footnote{MAP homepage: http://map.gsfc.nasa.gov}
and PLANCK\footnote{PLANCK homepage: http://astro.estec.esa.nl/Planck}, 
destriping techniques have already been  set up to clean the Time
Ordered Data (TOD) from low
frequency noise. The impact of different scanning strategies
on the quality of final maps has been studied as well (Delabrouille 1998; 
Maino et al. 1999).\\ 
The problem of removing $1/f$ noise from the data stream
is even more important for experiments attempting measurements of 
CMB polarization (CMBP), where even a tiny amount of 
residual correlated  noise, at a 
level negligible for investigations of CMB temperature anysotropy 
($\Delta T/T\simeq 10^{-5}$), might distort the signal characteristics: 
in the most favourable case the expected polarization level is 
only in the order of 10\% of the 
temperature fluctuations. The first algorithm specifically designed
for destriping polarization data has been proposed by 
Revenu et al. (2000), who extended to polarization data ideas previously
studied to destripe PLANCK anisotropy data (Delabrouille 1998).  \\
\begin{figure}
\includegraphics[width=0.65\hsize.,angle=90]{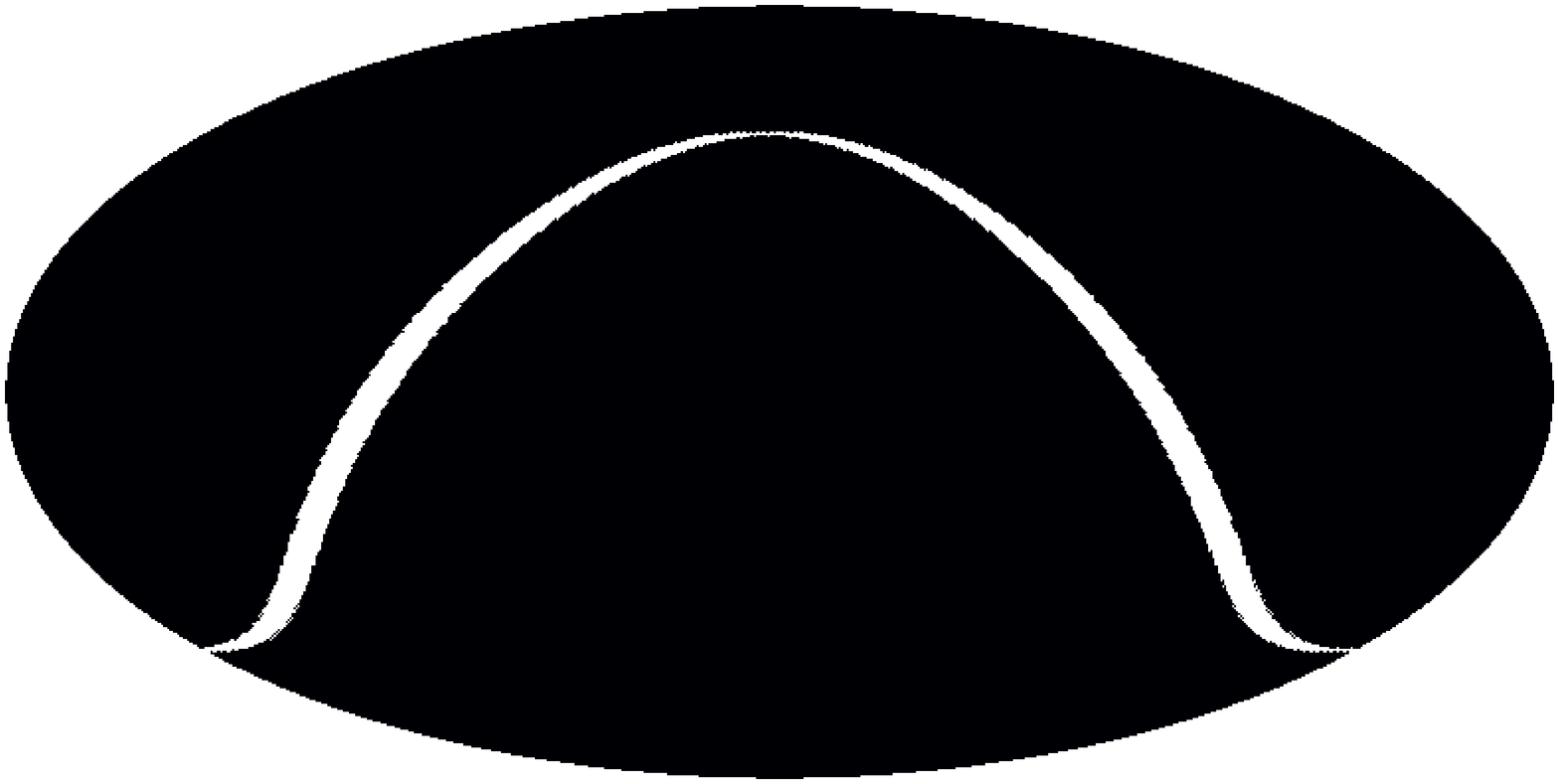}
\includegraphics[width=0.65\hsize,angle=90]{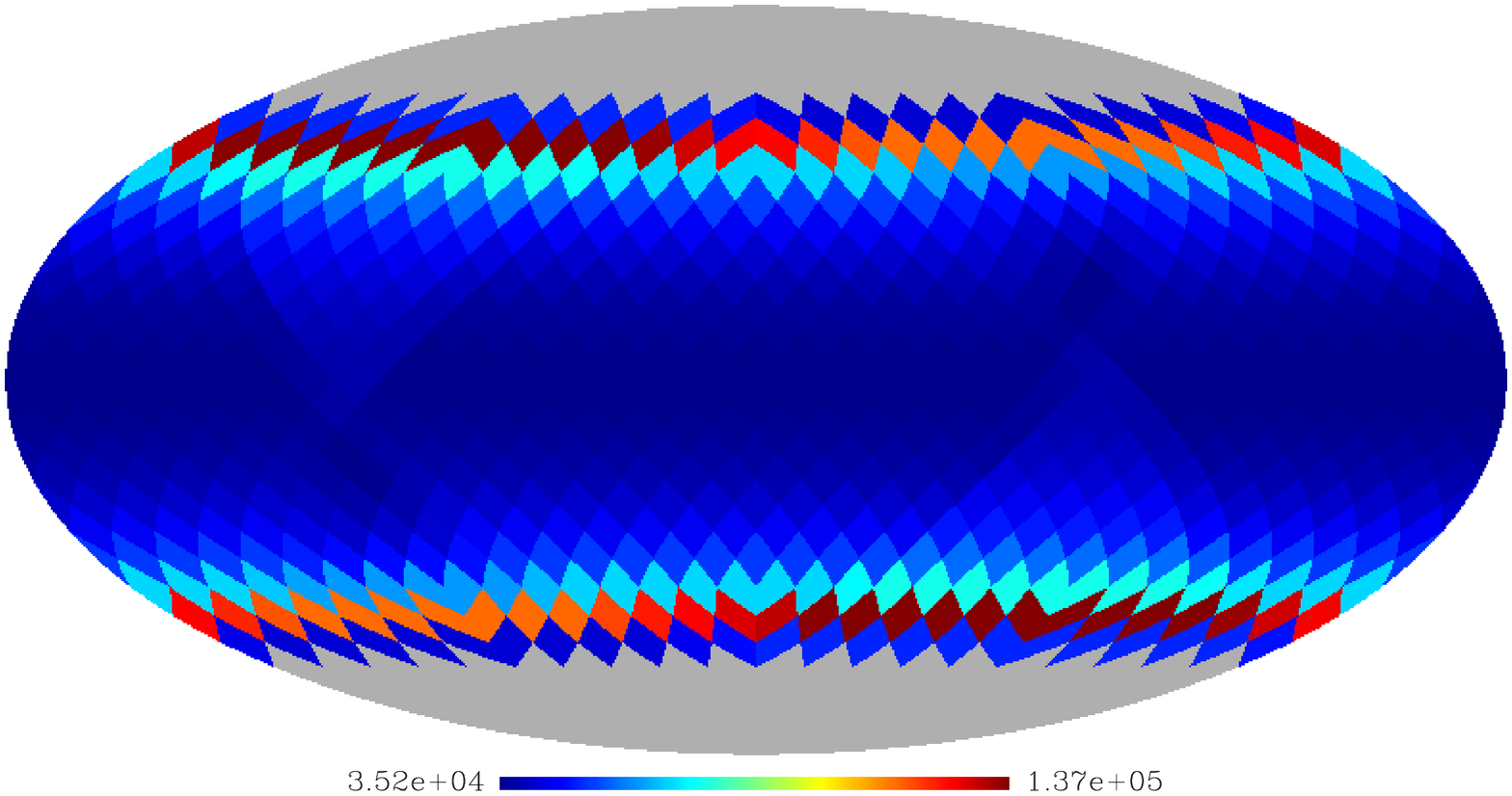}
\caption{Top: The sky, in Celestial coordinates, as scanned by SPOrt in
few orbits. Bottom: 
Pixel observing time (seconds) for 388 days of data taking. 
The pixel size is about $7^{\circ}$
 (HEALPix parameter $Nside=8$).} 
\label{fig:ISSorbit}
\end{figure}
The SPOrt\footnote{SPOrt homepage: http://sport.bo.iasf.cnr.it} experiment 
onboard the ISS (Cortiglioni et al. 2002; Carretti et al. 2003) 
is equipped with 4 correlation 
polarimeters directly measuring the 
$Q$ and $U$ Stokes parameters in the 22-90~GHz frequency range, with  
${\rm FWHM}=7^{\circ}$. 
Besides providing polarization maps of Galactic 
synchrotron, SPOrt
will try a first detection of CMBP on large angular scales 
($\ell\leq 25$), 
where the CMB E-mode power spectrum is particularly sensitive to the value 
of the optical depth of the reionized medium 
(Seljak 1997; Zaldarriaga et al. 1997). Such information cannot be 
extracted from CMB temperature anysotropy data, nor from CMBP data taken by
balloon and ground-based experiments looking at small sky patches.
This makes CMBP 
measurements on large angular scale of utmost importance. 
 
The algorithm presented here is particularly suited to SPOrt,
acting directely on $Q$ and $U$ Stokes parameter data. 

The SPOrt radiometers are
based on InP HEMT amplifiers, for which the low frequency noise
is known to have a $1/f$-like spectrum ($\beta\simeq 1$) dominated 
by transistor gain fluctuations (Wollack 1995), with knee frequencies 
in the range 100-1000~Hz. However,
when using correlation techniques, the knee frequency is reduced 
 by a factor $(T_{\rm offset}/T_{\rm sys})^2$, with $T_{\rm offset}$ 
and $T_{\rm sys}$ the instrumental offset and the system noise temperature, 
respectively (Wollack \& Pospiesalski 1998; 
Carretti et al. 2001). The present state of the instrument already guarantees
a value $f_{\rm k}<2.5\times 10^{-5}$~Hz (Carretti et al. 2003),
close to the goal knee frequency of one tenth of the ISS 
orbit frequency, i.e. $f_{\rm k}^{\rm goal}=1.8\times 10^{-5}$~Hz. \\
The SPOrt scanning strategy is bound to the ISS motion. The ISS
orbit is tilted by $51.6^{\circ}$ with respect to the 
Celestial equator and is characterized by a frequency 
$f_{\rm o}=1.8\times 10^{-4}$~Hz. 
The SPOrt antennae, looking at the zenit, will cover about 80\% of the 
sky under precessing circles intersecting each other. The nominal sampling 
rate is 1~Hz, and
a sky map will be provided every 70 days in a  
mission lasting at least 18 months.  
Further details about the SPOrt sky coverage  are shown in 
Fig.~\ref{fig:ISSorbit}.\\ 
%
The iterative algorithm described in this paper has
been specifically designed to destripe $Q$ and $U$ 
Stokes parameter data. 
It is
general enough to be valid for any experiment where data are taken by
scanning the sky at a frequency higher than the receiver knee frequency, 
and where
different scans intersect each other in a sufficient number of points. 
Measurements of scalar quantities are easily handled. \\
The description of the algorithm is provided in Sect.~\ref{sec:method}, 
whereas Sect.~\ref{sec:tests}
presents various
tests of its performances and Sect.~\ref{sec:conclusions} summarises our
conclusions.

\section{Destriping method}
\label{sec:method}
The only assumptions our algorithm relies upon is that the radiometer 
is stable during the 
signal modulation period (the time needed to complete one orbit in case 
of SPOrt), so that the
instrumental offset does not change significantly during this time, 
and there is enough overlap between different orbits. 
In such a case the noise
can be split in two parts: for time scales shorter than the orbit 
period (high frequencies) it is essentially white, whereas for longer timescales 
(low frequencies) it also contains the 
$1/f$ component, which we model as a different constant offset for 
each orbit.
A simple iterative procedure is then   
applied to remove these offsets from the TOD before 
map-making. 
The map-making itself consists in averaging the cleaned TOD  corresponding 
to the same sky 
pixel after rotating them into a fixed reference frame. Parallel transport 
to the center of each pixel 
is also implemented, as described by Bruscoli et al. (2002), before averaging, 
and can optionally be
switched on, though it is not strictly necessary
for the portion of sky covered by SPOrt. It would be mandatory if polar 
caps were included in the
accessible region. The time needed to simulate 1 year of SPOrt data taking
with the nominal sampling rate of 1~Hz,
destripe the TOD and produce
$Q$ and $U$ clean maps is about $10$ minutes  
on a 1.3 GHz Pentium III equipped with a 4~GB~RAM
when using 10 destriping cycles, and increases by 2.5 minutes for each
additional bunch of 10 iteration cycles.

\subsection{Formalism}
Each datum consists of a ($\tilde{Q}$,$\tilde{U}$) pair as measured 
in the polarimeter 
reference frame: the same sky point, when seen from different orbits, shows 
different ($\tilde{Q}$,$\tilde{U}$)
values  even in absence of noise. The standard longitude-latitude 
fixed reference frame (Berkhuijsen 1975) is used 
when considering the real sky emission, which is projected into an $N$-pixel 
map.
Our $M$ TOD, collected during $R$ orbits about 
the Earth, are used to build an $M$-vector \Y\ 
of measured ($\tilde{Q}$,$\tilde{U}$) pairs:
\begin{equation}
\Y= \A\cdot\X+\B\cdot\Off+\Nn.
\label{eq:datamodel}
\end{equation}
\X\ is an $N$-vector whose elements are the ($Q$,$U$) pairs of the true 
sky emission;  \A\ is an $M \times N$ 
pointing matrix  connecting sky pixels to 
observations (Wright 1996; Tegmark 1997a): its elements are $2\times2$
matrices rotating 
the Stokes parameters from the fixed to the polarimeter 
reference frame;
\Off\ is an $R$-vector containing a pair of radiometric offsets 
for each orbit;
\B\ is an $M \times R$ pointing matrix connecting 
orbits to observations, 
and \Nn\ is an $M$-vector containing  pairs of white noise.
In more details, the elements of the above mentioned vectors and 
matrices are:
 \begin{eqnarray}	
 Y_t =  
 \left(
      \begin{array}{c}
      \tilde{Q}_t\\
      \tilde{U}_t
      \end{array}
 \right)
\end{eqnarray}	
\begin{equation}
    A_{tp} = \left\{
              \begin{array}{ll}
              {\mathcal R}(\alpha_{t}) =
              \left(
                      \begin{array}{cc}
                      \cos (2\alpha_{t}) & \sin (2\alpha_{t})\\
                      -\sin (2\alpha_{t}) & \cos (2\alpha_{t})
                      \end{array}
              \right)
              & \begin{array}{l}
                       \,\,\,\!\!\!\mbox{if pixel } p \\
                      \,\,\,\!\!\!\mbox{is observed}\\
                      \,\,\,\!\!\!\mbox{at time } t
                \end{array}\\
                \\
              \left(
                      \begin{array}{cc}
                      0 & 0\\
                      0 & 0
                      \end{array}
              \right)
                & \,\,\,{\rm otherwise}
              \end{array}
              \right.
 \end{equation}
where $\alpha_{t}$ is the angle between the polarimeter and the
fixed reference frames at time $t$.
\begin{eqnarray}	
 X_p =
 \left(
      \begin{array}{c}
      Q_p\\
      U_p
      \end{array}
 \right)
\end{eqnarray}	
\begin{equation}
    B_{to} = \left\{
              \begin{array}{ll}
                 \left(
                    \begin{array}{cc}
                    1 & 0\\
                    0 & 1
                    \end{array}
                 \right) & \,\,\,\,\mbox{if } o \mbox{ is the orbit
                  run at time } t\\
                 \\
                 \left(
                 \begin{array}{cc}
                 0& 0 \\
                 0 &0
                 \end{array}
                 \right) & \,\,\,\,{\rm otherwise}
              \end{array}
              \right.
\end{equation}
\begin{eqnarray}	
 O\!f\!f_o = 
 \left(
     \begin{array}{c}
         O\!f\!f^{\,\tilde{Q}}_o\\
         O\!f\!f^{\,\tilde{U}}_o
      \end{array}
 \right)
 \end{eqnarray}
 \begin{eqnarray}	
 N_t =
 \left(
      \begin{array}{c}
      N^{\tilde{Q}}_t\\
      N^{\tilde{U}}_t
      \end{array}
 \right)
\end{eqnarray}

\subsection{Iteration cycle}

If we knew the sky emission vector \X\  we could use it to obtain the offset 
vector \Off\ 
by inverting Eq.~(\ref{eq:datamodel}). 
Of course, estimating the noise offsets would not  be worthwhile if we 
already knew the true sky
emission. However, to start we can guess a 0-order sky emission 
${\bf X^{(0)}}$,
the most neutral choice being a null vector; then obtain a 0-order offset 
estimate, and subtract it from our measurements to get a 0-order
cleaned TOD; from this compute ${\bf X^{(1)}}$, the 1$^{\rm st}$-order 
estimate of the real sky emission, and iterate the cycle.\\
Starting from Eq.~(\ref{eq:datamodel}) and
following Tegmark (1997b), the best 
estimators  for the offset vector and the signal matrix 
can be written:
\begin{eqnarray}
\Off& = & \Wos(\Y - \A\cdot\X)\\
\X& = &\Wpix(\Y - \B\cdot\Off)          
\end{eqnarray}
where, in case of uniform white-noise variance, as
expected for the SPOrt receivers,
\begin{eqnarray}
\Wos& = & \Nos^{-1}\cdot\B^{\rm T}\\
\Wpix& = &\Npix^{-1}\cdot\A^{\rm T}         
\end{eqnarray}
with $\Nos=\B^{\rm T}\B$ and $\Npix = \A^{\rm T}\A$ diagonal 
matrices containing the number of observations in each orbit  
and in each pixel, respectively. In such a case
Eq~(10) simply corresponds to
 rotating  into the fixed reference frame
 the offset-subtracted observations and averaging those 
corresponding to the same sky pixel.
When starting from a null guess map, i.e. $\X^{(0)}=0$, the first offset
 estimate is a simple  
average of the measurements taken over each orbit.

In the more general case of non-uniform white noise, i.e. 
when the noise covariance
matrix $\Cc=\langle\Nn\Nn^{\rm T}\rangle$ has unequal diagonal elements, 
the matrices $\Wos$ and $\Wpix$  become (Tegmark 1997b)
\begin{eqnarray}
\Wos& = & (\B^{\rm T}\Cc^{-1}\B)^{-1}\B^{\rm T}\Cc^{-1}\\
\Wpix& = &(\A^{\rm T}\Cc^{-1}\A)^{-1}\A^{\rm T}\Cc^{-1}
\end{eqnarray}
meaning that the  simple averages mentioned above should be
replaced with noise-weighted averages. 

In any case the first offset estimate writes:
\begin{eqnarray}
\Off^{(0)} & = & \Wos^{-1}\Y
\end{eqnarray}
 The 0-order cleaned TOD are:
\begin{equation}
\Y^{(0)}  =  \Y - \B\cdot\Off^{(0)} 
\end{equation}
and the 1$^{\rm st}$-order estimate 
of the true sky emission  is
\begin{equation}
\X^{(1)}=\Wpix\cdot\Y^{(0)}
\end{equation}
The $i^{th}$ cycle writes:
\begin{eqnarray}
\Off^{(i)}& = & \Wos(\Y - \A\cdot\X^{(i)})\\
\X^{(i+1)}& = &\Wpix (\Y - \B\cdot\Off^{(i)})          
\end{eqnarray}
and must be iterated
until the difference between the sky emission estimates of order $i$ 
and $i+1$ is negligible
when compared to the white noise level.

The proof that the procedure is effective has been obtained numerically, 
as described in detail in the next section. Destriping measurements of total 
intensity (or, in general, any other
scalar quantity) can be simply 
performed by replacing pairs with scalars in the 
previously 
defined vectors, and rotation matrices with scalar 1. 
The average sky signal is lost when destriping maps
of scalar quantities. However, for maps of polarization data like $Q$ and $U$
it can be kept provided the polarimeter
reference frame rotates about the fixed frame while running along each
orbit, as in the case of SPOrt. This is a nice feature, especially 
when measuring foreground contributions.

\section{Simulations and tests}
\label{sec:tests}

We test the method with numerical simulations of the data
stream expected from about one year of SPOrt data taking. \\ 
As real sky emission we consider both CMB and synchrotron radiation, both 
convolved with a $7^{\circ}$ Gaussian beam. Other foregrounds are 
expected to be less important at the frequencies and 
angular scales covered by SPOrt (Tegmark et al. 2000; Lazarian 
\& Prunet 2002) and are not included.\\
CMB emission is generated by the 
CMBFAST\footnote{http://www.physics.nyu.edu/matiasz/CMBFAST/cmbfast.html} 
package
according to a $\Lambda$CDM cosmological model
with $\Omega_m=0.3$, $\Omega_{\Lambda}=0.7$, $\tau=0.2$, 
$H_0=65~{\rm Km}{\rm s}^{-1}{\rm Mpc}^{-1}$,
$\Omega_{b}=0.05$ and no contributions from gravitational waves. 
For the synchrotron radiation
 we use $Q$ and $U$ map templates at 22~GHz
developed by Bernardi et al. (2003) and featuring a 
$P_{\rm rms}=\sqrt{\langle Q^2+U^2 \rangle}\simeq 35~\mu$K.
The 
polarized intensity 
peak emission 
is about 130~$\mu$K.
The main ingredients of the template are total intensity low frequency data 
(Haslam et al. 1982; Reich 1982), 
from which the syncrothron polarized intensity is extracted, 
and optical starlight data (Heiles 2000), from which polarization 
angles are derived. 
The template frequency has been chosen to correspond to the 
SPOrt channel where synchrotron emission is expected to be most important.

Noise streams to be added to the sky signal are built in the frequency 
space, where their Fourier coefficients are generated 
according to the spectrum defined in Eq.~(\ref{eq:noiseps}), 
with $\beta=1$, 
and  transported to the time domain via FFT.
Our code allows us to calculate the angle between our polarimeters
and the longitude-latitude fixed reference frame for each data sample, and 
to set the sampling
frequency from the nominal rate (1~Hz) to any other rate while keeping the
SPOrt nominal istantaneous sensitivity of 
$1~{\rm mK}\,{\rm s}^{1/2}$~(Carretti et al. 2003). 
In the tests described here the sampling rate
was set to 1~Hz, and the SPOrt observing time was about 1 year. 
The final output of our simulations are TOD and noise arrays, 
clean $Q$ and $U$ signal maps to investigate the effects of SPOrt partial 
sky survey, $Q$ and $U$ maps containing both signal and noise before and after 
destriping as well as a hit map to evaluate the time spent over each 
map pixel.
We use the HEALPix\footnote{http://www.eso.org/science/healpix} 
pixelisation scheme (Gorski et al. 1998) 
 with $Nside=64$ to
ensure the effects of pixelisation are negligible when compared to the
antenna beam (the pixel size of the output maps is $\simeq 1^{\circ}$). \\
We test the destriping efficiency on simulated maps
by evaluating the residual low frequency noise
 in terms of added rms noise with respect to the white noise case,
 angular correlation functions $C^X(\theta)$ ($X=Q,U$) and 
power spectra $C^{Y}_{\ell}$ ($Y=E,B$).

\subsection{Correlation functions and power spectra}

To avoid edge effects due to partial sky coverage and to efficiently take
into account the presence of noise, even in case of non uniform sky coverage,
we primarly test our destriping technique by studying the two-point 
correlation functions $C^X(\theta)$ ($X=Q,U$) measured from our simulated 
maps. We sample the correlation functions with about one degree of angular 
resolution according to the expression:
\begin{equation}
\tilde{C}^X(\theta)=
\frac{\sum_{ij}w_{ij}\Delta^X_i\Delta^X_j}{\sum_{ij}w_{ij}}.
\end{equation}
Here $\Delta^X_i$ is the content of pixel $i$ of map $X$, 
and $w_{ij}$ is the weight 
of the $ij$ pixel pair
\begin{equation} 
w_{ij} =\frac{2w_iw_j}{w_i+w_j}
\end{equation} 
where each pixel weight $w_{i}$ is proportional to the pixel observing time 
and  $\sum_{i=1}^{N_{pix}}w_{i}=N_{pix}$.
\begin{figure}
\includegraphics[width=1.0\hsize]{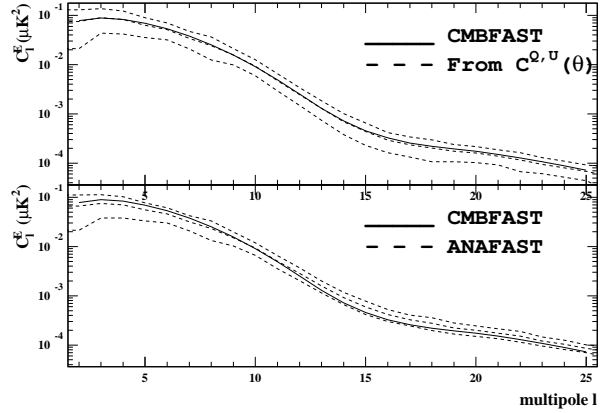}
\caption{$C^E_{\ell}$ power spectrum of our selected CMB model,
convolved with the SPOrt nominal FWHM of 7$^o$ (solid), compared to the 
average of 100 power spectra measured by either our method (top)
or ANAFAST (bottom) on simulated maps featuring the
SPOrt cuts. Measured 1-$\sigma$ bands due to cosmic variance are
shown as well.} 
\label{fig:psmethod}
\end{figure}
In the computation 
the Stokes parameters are
properly rotated
so that $Q$ is defined along the great circle 
connecting the points in each pair
(Zaldarriaga 1998). The exact value of the angle $\theta$
at which the correlation functions are
 sampled is calculated as the weighted average 
of the angles between pixels contributing to the same measured point.
As an example, the measured two-point correlation functions of $Q$,$U$ maps 
containing only white noise are on average zero everywhere 
but at 0-distance, where they are checked to 
be equivalent to the average pixel variance.\\
The $C_{\ell}^{E,B}$ 
power spectra can  be  recovered from the measured 
correlation functions by 
integration (Kamionkowski et al. 1997; Zaldarriaga 1998):
\begin{equation}
 C_{\ell}^{ E}=\int^{\pi}_0[C^{ Q}(\theta)F^1_{\ell,2}(\theta)+
C^{U}(\theta)F^2_{\ell,2}(\theta)]\sin\theta d\theta
\end{equation}
\begin{equation}
 C_{\ell}^{ B}=\int^{\pi}_0[C^{ U}(\theta)F^1_{\ell,2}(\theta)+
C^{ Q}(\theta)F^2_{\ell,2}(\theta)]\sin\theta d\theta
\end{equation}
where 
the functions $F^i_{\ell,2}(\theta)$ have been defined by Zaldarriaga (1998).
We perform the integration by Gauss quadrature, where 
the correlation function values at the needed points $\theta$
are obtained by interpolation, with a second order polynomial, of the 
three closest measured points.  
This method of measuring power spectra is similar to that used 
by Szapudi et al. (2001), though 
here it has been extended to polarization data. 
It has been tested on full-sky maps of simulated CMB emission, convolved
with a Gaussian filter with FWHM=7$^{\circ}$, by comparing the measured 
power spectra to those obtained by using the ANAFAST package: we get 
the same results,  within numerical precision, up to $\ell=25$, 
i.e. in the full multipole range covered by SPOrt. 
On partial sky maps this method is found
to work better than ANAFAST, thanks to the lack of edge effects:
our average of power spectra 
extracted from CMB maps without noise and featuring the SPOrt
sky coverage matches the input theoretical spectrum very well,
whereas that measured with ANAFAST is flatter, as shown 
in Fig.~\ref{fig:psmethod}. Furthermore,
when trying to extract the real sky signal from noisy maps
this method allows for pixel weighting in case of non-uniform sky
coverage as well as noise subtraction, provided the noise
statistical properties (correlation functions) are known. 
It has the additional advantages of 
being simple and rather fast:
the required computing time practically corresponds to that
needed to calculate the two-point correlation functions, i.e. about 35
minutes on our platform. 
The number of operations is $O(N_{\rm pix}^2)$, while optimal methods 
to measure power spectra require matrix inversions tipically 
implying a number of  operations $O(N_{\rm pix}^3)$ (Wright 1996). 
The implementation of a new algorithm for
computing N-point correlation functions is expected to further reduce the
computing time (Moore et al. 2000). \\

\subsection{Destriping signal-only maps}

To check for the presence of spurious correlations introduced by our algorithm
and depending on the underlying signal,
as a first test we run the destriping code over CMB and Synchrotron 
maps containing no
noise. The test confirms our algorithm subtracts the average signal 
from the input map when dealing with scalar quantities 
(CMB temperature  in our case), whereas
nothing is lost in our $Q$ and $U$ maps.
We then analyse the difference between the destriped and the original maps
as a function of the number of iterations of the destriping cycle. 
Should no residual
effects be present, these differences would asimptotically 
become null for $Q$ and $U$ maps, and a 
constant number corresponding to the average of 
the input signal for scalar maps. 
Indeed, the maximum pixel-to-pixel difference, 
which depends on the level of the underlying signal, decreases 
with increasing number of iterations.
In particular, the spurious noise introduced when destriping maps of 
synchrotron emission at 22~GHz, i.e. the SPOrt frequency where 
the Galactic contribution is expected to be higher,  becomes negligible 
also for CMBP after about 20 loops, reaching the tiny level of
few nK after 30 loops. More details are shown in 
Table~1, both in terms of
peak-to-peak amplitude and rms of spurious noise. For comparison, the 
peak-to-peak amplitude and rms of the
underlying signal maps are shown as well.  
\begin{table}[htb]
\label{tab:spurnoise}
\caption{Peak-to-peak amplitude and rms 
of the spurious noise introduced when destriping signal-only maps (CMB 
and synchrotron), as functions of the  
iteration loop number, for $Q$ and $U$ maps.
The last line shows the peak-to-peak amplitude and rms of the underlying 
$Q$~\&~$U$ signals.}
\begin{center}
\begin{tabular}{lllll}
\hline
 {\bf N$_{\rm loops}$} &  \multicolumn{2}{c}{\bf cmb}   & 
\multicolumn{2}{c}{\bf sync } \\
\hline
 & peak-to-peak &  rms &  peak-to-peak &  rms \\
 & amplitude [$\mu$K] & [$\mu$K]  & amplitude [$\mu$K] &  [$\mu$K] \\
\hline
10 &  $3\times10^{-2}$ & $4\times10^{-3}$ &  1.1 
&              $ 0.3$           \\
20 &  $2\times10^{-3}$ & $4\times10^{-4}$ &  $8\times10^{-2}$ 
& $2\times10^{-2}$\\
30 &  $3\times10^{-4}$ & $3\times10^{-5}$ &  $1\times10^{-2}$ 
& $2\times10^{-3}$\\
40 &  $1\times10^{-4}$ & $4\times10^{-6}$ &  $5\times10^{-3}$ 
& $2\times10^{-4}$\\\hline
   & 3 & 0.5 & 170 & 20 \\ \hline 
\end{tabular}
\end{center}
\end{table}  

The same conclusions can be derived by
inspecting the angular correlation functions  $C^X(\theta)$ ($X=Q,U$) of 
the spurious noise: they are not flat, but
deviations from a flat null function decrease with increasing number of 
iteration cycles and can be made negligible after
about 10 (20) loops when destriping CMB (synchrotron) $Q$, $U$ maps.

\subsection{Destriping noise maps }

\begin{figure}
\label{fig:mapdes}
\includegraphics[width=0.65\hsize,angle=90]{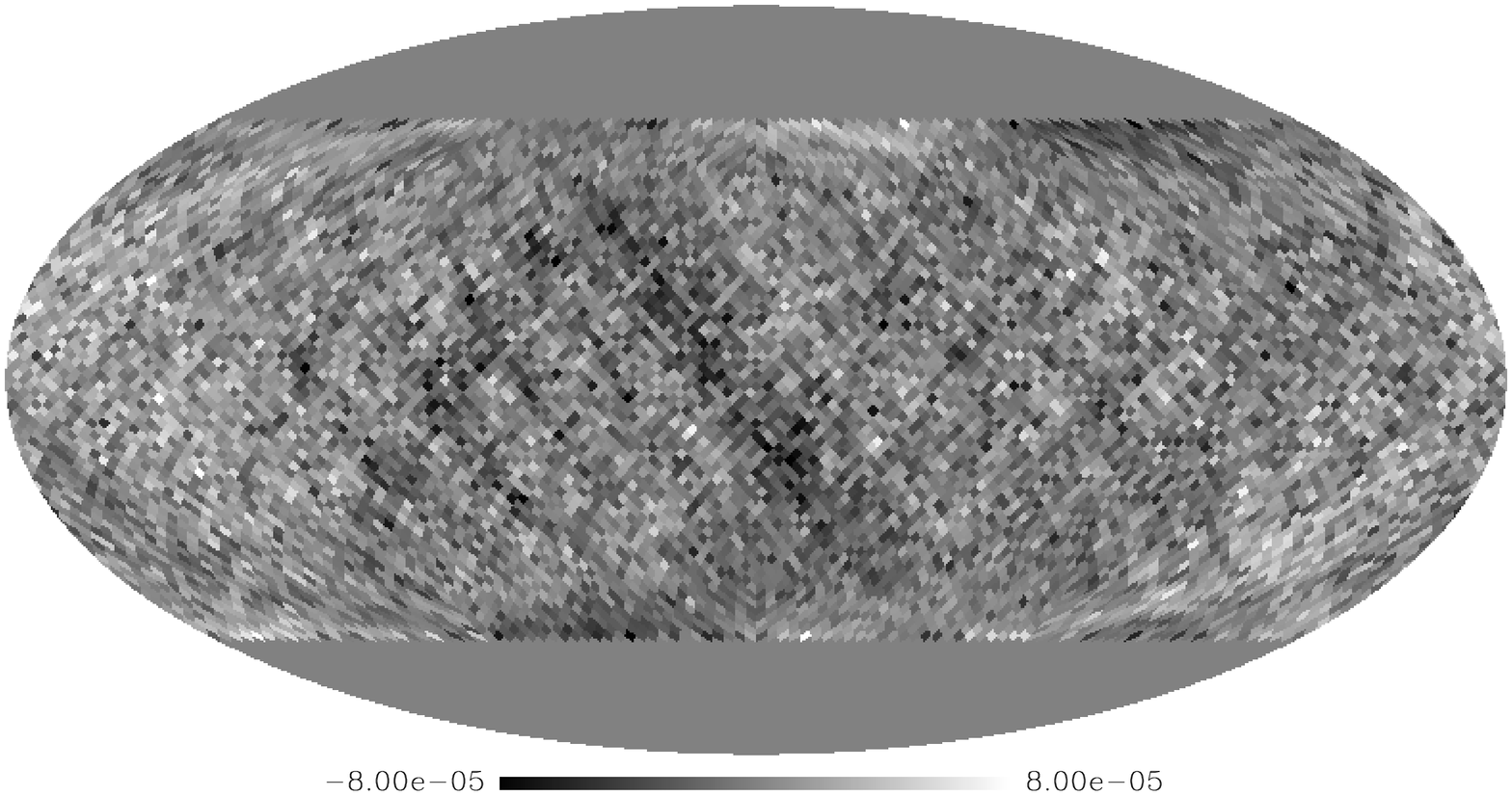}
\includegraphics[width=0.65\hsize,angle=90]{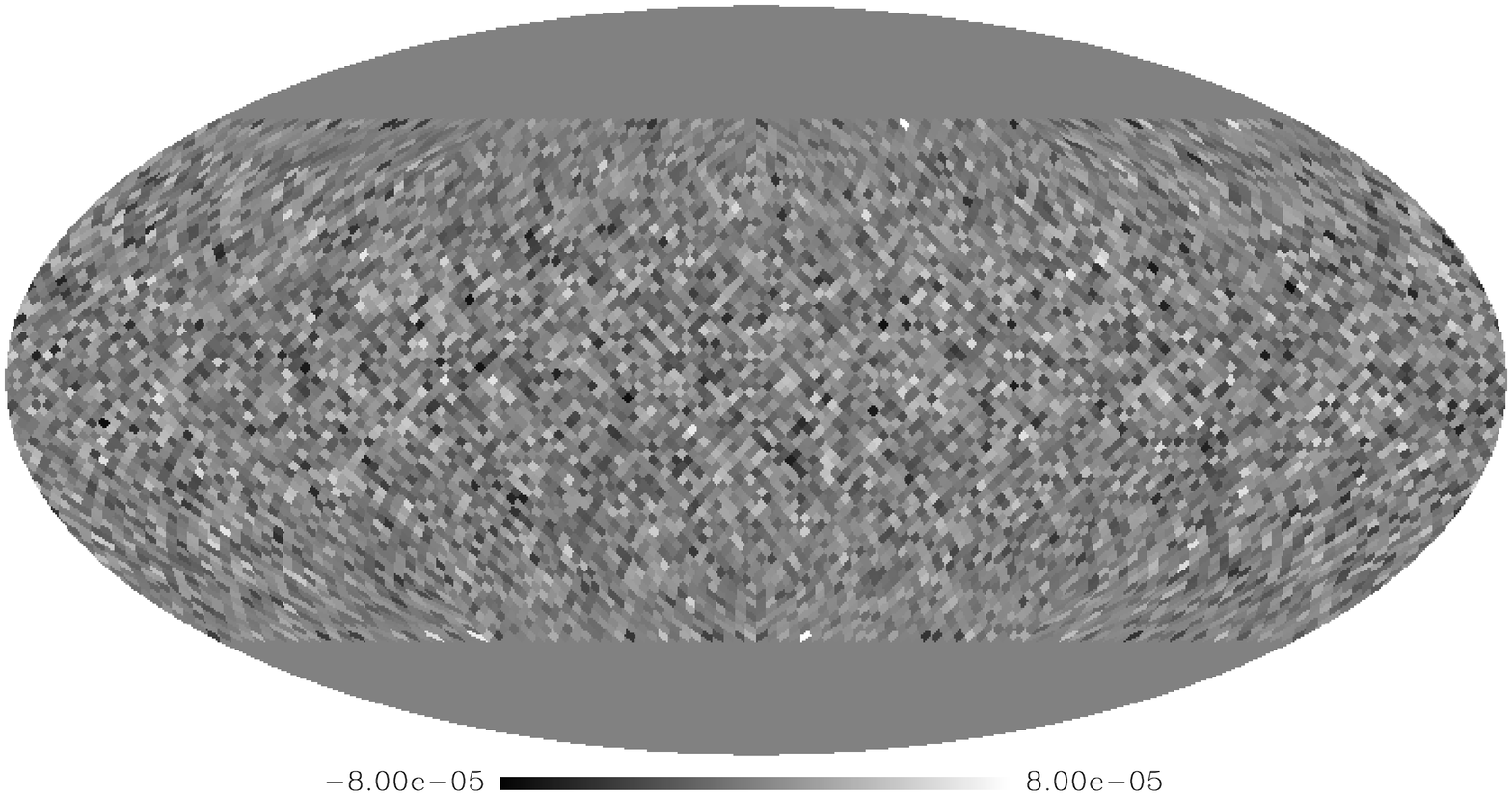}
\caption{Noise simulated maps before (top) and after (bottom) destriping, for 
the case $f_k=1.8\cdot 10^{-4}$, in Kelvin. The HEALPix parameter
$Nside$ is 32.} 
\label{fig:mapdesp}
\end{figure}

The effectiveness of our destriping technique is then tested on simulated maps 
containing only noise, in the following cases:
\begin{itemize}
  \item{} white noise 
  \item{} $1/f$ noise
  \item{} both white and $1/f$ noise
\end{itemize} 
The good performances of the technique are made evident
in Fig.~\ref{fig:mapdesp} where we show a simulated noise
map (both $1/f$ and white) before and after destriping.

One possible way to quantify the residual correlated noise 
after destriping is measuring 
the fractional excess pixel noise with respect to the 
case of purely white noise. Results are shown in Table~2 
for two different values of
the knee frequency, corresponding to the SPOrt goal knee frequency, 
$f_{\rm k}=1.8\times 10^{-5}$~Hz, and the SPOrt orbit frequency,
$f_{\rm o}=1.8\times 10^{-4}$~Hz, 
the latter representing a very conservative case.
\begin{table}[htb]
\label{tab:rmsnoise}
\caption{
Excess rms noise due to low-frequency contributions,
with respect to the white noise level, for pixels of $\simeq 7^{\circ}$.}
\begin{center}
\begin{tabular}{lll}
\hline
$\mathbf{f_{\rm k}}$~{\bf(Hz)} & {\bf Before Destriping} & {\bf After Destriping} \\
\hline
$1.8\times 10^{-4}$ & 310\% & 6\% \\
$1.8\times 10^{-5}$ & 35\% & $<1$\% \\ \hline
\end{tabular}
\end{center}
\end{table}

\begin{figure*}
\centering
\includegraphics[width=1.05\hsize]{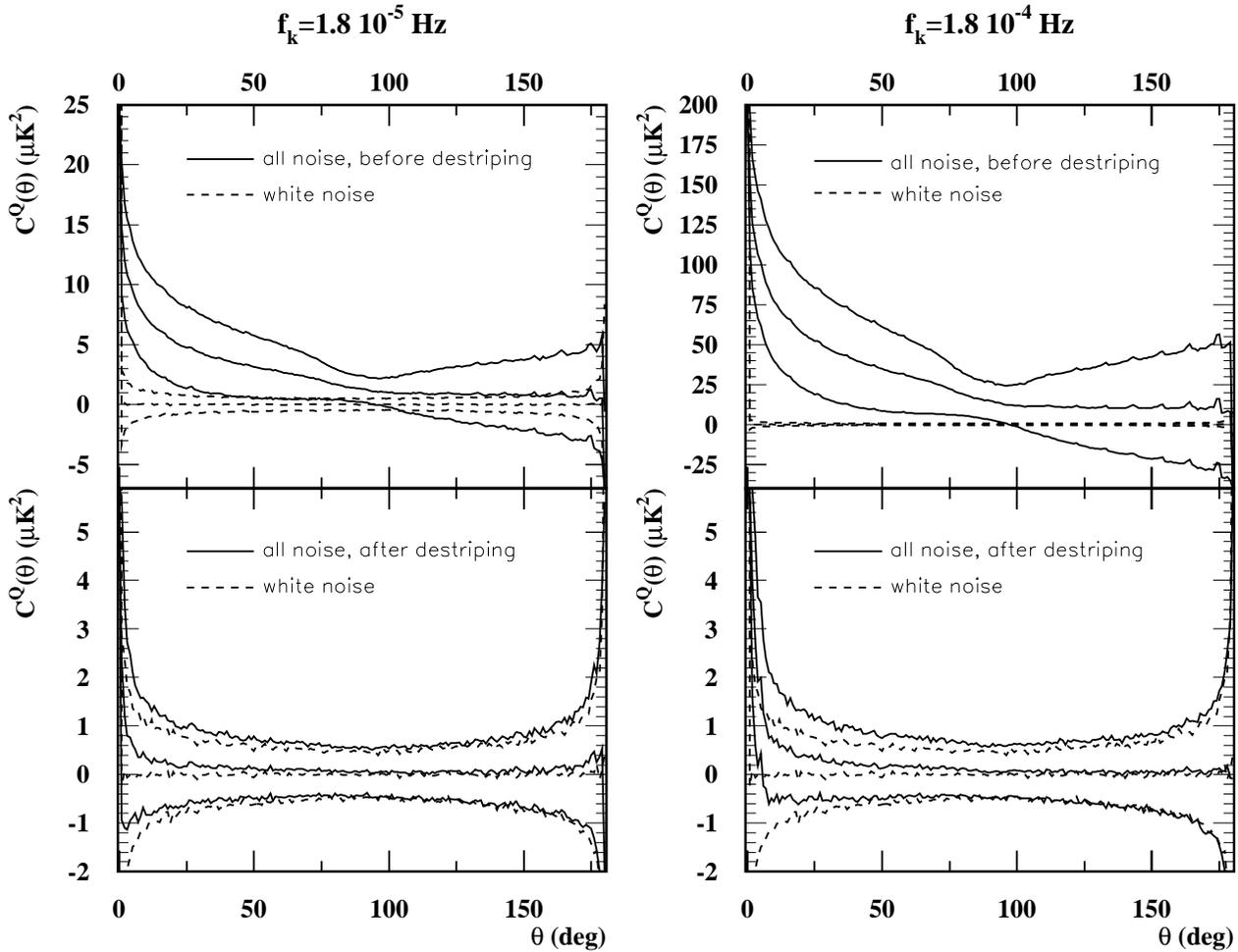}
\caption{Average and 1$\sigma$ band of  correlation functions $C^Q(\theta)$
 measured from 500 noise maps,  
before and after destriping, for two
values of the knee frequency $f_{\rm k}$.  
Before destriping the $y$ scale is not the same for the 
two $f_{\rm k}$ values.
Each simulation
corresponds to about one year of SPOrt realistic data taking. 
The case of purely white noise is shown for comparison.}
\label{fig:noisecorr}
\end{figure*}

Another way to quantify the residual correlated noise is 
measuring and inspecting the two-point correlation 
functions C$^X(\theta)$ of 
simulated $Q$ and $U$ noise maps. 
Averages and 1$\sigma$ bands of 500 correlation functions 
C$^{Q}(\theta)$
measured from maps containing both white and $1/f$ noise, before and 
after destriping, are compared
to the purely white noise case in Fig.~\ref{fig:noisecorr} for the 
same knee frequencies as in the previous test. 
The correlation functions
C$^{U}(\theta)$  are similar and are not shown.
As expected, the correlated noise is strongly reduced by the destriping 
procedure, the residuals falling within the statistical error of
the white noise case for  $f_{\rm k}=f_{\rm k}^{\rm goal}$.
The correlation functions for the case of purely $1/f$ noise are similar 
to the white plus $1/f$ case everywhere but at 0 distance, and 
are not shown.

If the noise statistical properties are known, its two-point correlation
function after destriping
can be calculated with Monte Carlo techniques and subtracted
from the measured values before integrating, the recovering of 
signal power spectra presenting no huge problems.
However, the residual correlated noise also increases
the expected error on measured quantities. In our case the rms
of  correlation functions  measured from many noise maps  
is increased by roughly  4\% (11\%),  for $f_{\rm k}=1.8\times 10^{-5}$~Hz 
($f_{\rm k}=1.8\times 10^{-4}$~Hz),
with respect to the case of purely
white noise. 

\begin{figure*}
\centering
\includegraphics[width=1.05\hsize]{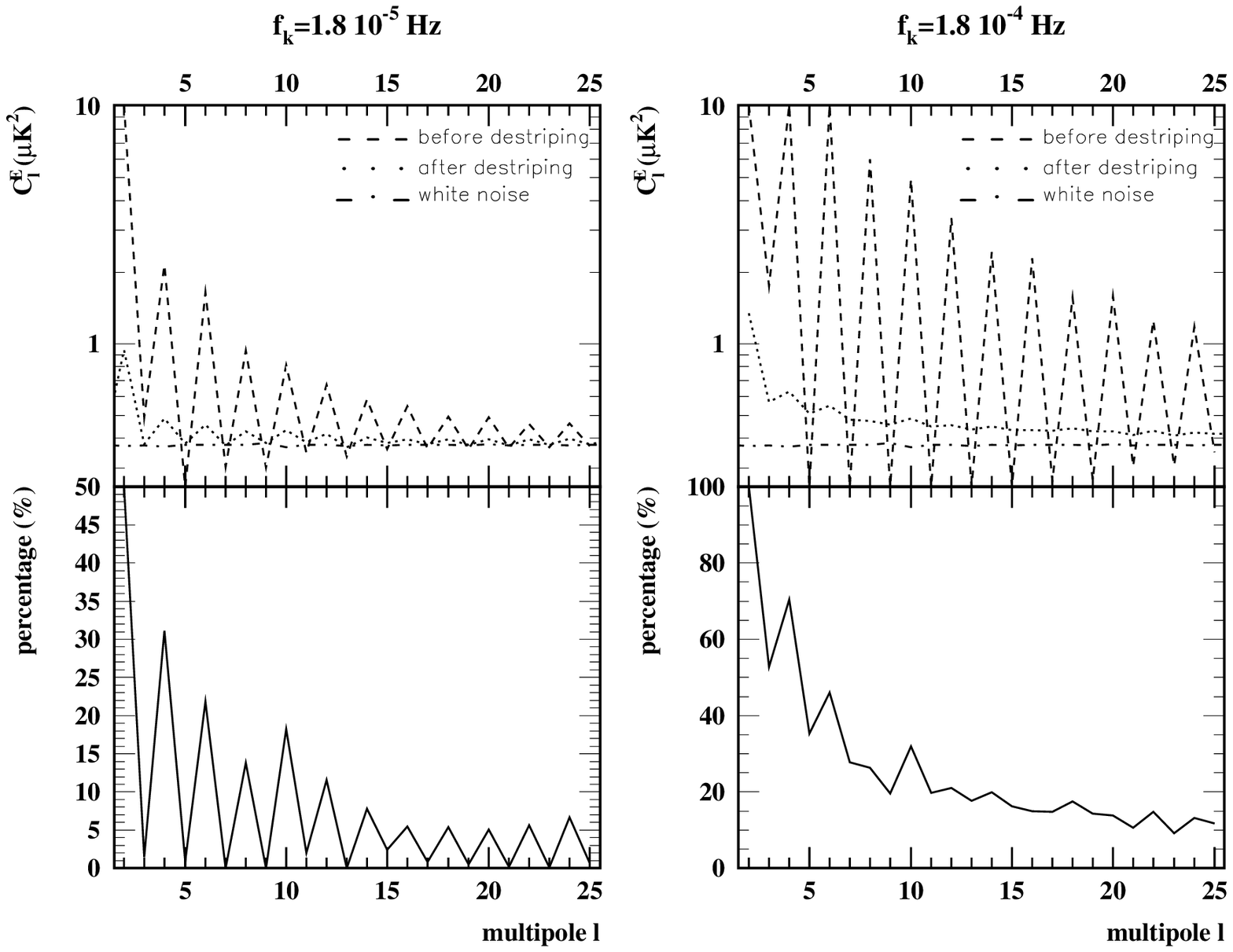}
\caption{Average of noise power spectra measured from the same maps
used to make Fig.~\ref{fig:noisecorr} (top). 
In the bottom panel the residual noise
in percentage of the measured white noise level is shown as a 
function of the multipole.}  
\label{fig:nps}
\end{figure*}
\begin{figure}
\includegraphics[width=1.0\hsize]{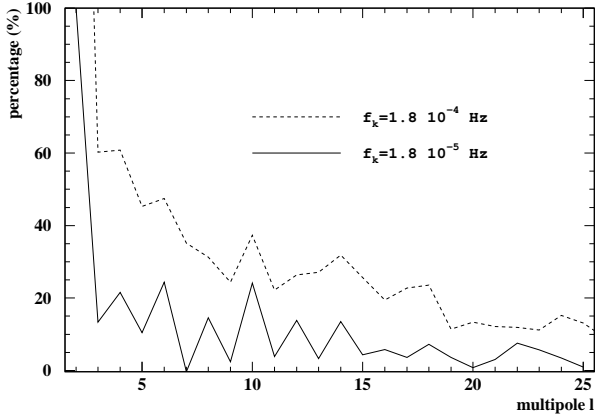}
\caption{Increment, due to the presence
of residual correlated noise, in the  rms of measured 
noise power spectra, after destriping, as a function of 
the multipole, in percentage of the rms of purely
white-noise power spectra. }  
\label{fig:clserror}
\end{figure}
For completeness, and to ease the comparison with other 
methods, average noise power spectra, obtained from the correlation functions
 used to make Fig.~\ref{fig:noisecorr}, are   
shown in Fig.~\ref{fig:nps}, though they carry the same information
as the correlation functions.
As expected, even for the largest knee frequency 
the $C^E_{\ell}$ power spectrum 
of the residual noise is close to that of purely white noise.
The region of low multipoles is the most sensitive to low frequency 
 residuals, some
contributions being always found here also after the application of 
other destriping 
techniques (Maino et al. 1999). 
The excess noise, in percentage of the measured white noise level,
is shown in the bottom panel of Fig.~\ref{fig:nps} as a function 
of the multipole.
The theoretical white noise level can be calculated as follows:
\begin{equation}
C_{\ell}^{wn}=\frac{4\pi}{N_{pix}}\langle \sigma^2 \rangle
\end{equation}
where $\langle \sigma^2 \rangle$ represents the average noise variance
per pixel.
Our measured value is equivalent to its 
expectation of $0.37~\mu$K$^2$, again confirming that our map-making
algorithm is correct. 

Finally, Fig.~\ref{fig:clserror} shows the increment, 
due to the presence
of residual correlated noise, in the rms of measured noise power 
spectra, after destriping, in percentage of the rms of purely
white noise power spectra, again as a function of the multipole. 
                                        
\section{Conclusions}
\label{sec:conclusions}
We presented a new iterative destriping technique working directely on
$Q$ and $U$ Stokes parameter data as well as 
on scalar quantities. As for other methods, in order for the
technique to work properly the receiver knee frequency must be
lower than the experiment modulation frequency, and there must be
sufficient overlap between different orbits. On the other hand,
no requirements on the statistical properties of the noise are
necessary here. Moreover,
our observing strategy preserves average $Q$ and $U$ values
 in the destriped maps, though 
the map average is still lost when dealing with scalar quantities. 

The performances of the technique, 
studied on simulated SPOrt data by analysing both the measured 
$C^{Q,U}(\theta)$ two-point correlation functions and 
the measured $C^{E,B}_{\ell}$ power
spectra, are comparable to those of other methods
based on $\chi^2$ minimisation. 
Computer time is not an issue, no matrix 
inversions being involved.
 
Power spectra are 
recovered from the measured two-point correlation functions 
by integration, this method being implemented here 
for the first time for polarization data. 
It has the advantages of
avoiding edge problems arising when using the ANAFAST code
and to allow for pixel weighting in case of 
non uniform sky coverage.

\begin{acknowledgements}
  This work has been carried out in the context of
  the SPOrt program, which is funded by the Italian Space Agency.
  We thank B.~Audone and F.~Amisano for useful discussions
  and the anonimous referee for good suggestions.
  We acknowledge the use of CMBFAST and HEALPix packages.
\end{acknowledgements}

\end{document}